# Enhanced Denoising of Optical Coherence Tomography Images Using Residual U-Net

Akkidas Noel Prakash[a], Jahnavi Sai Ganta[a], Ramaswami Krishnadas[b], Tin A. Tun[c], Tin Aung[c,d,e], Satish K Panda[a,f]

[a] School of Mechanical Sciences, IIT Bhubaneswar, India

[b] Glaucoma Services, Aravind Eye Care Systems, Madurai, India.

[c] Singapore Eye Research Institute, Singapore

[d] Duke-NUS Medical School, Singapore

[e] Department of Ophthalmology, Yong Loo Lin School of Medicine, National University of Singapore, Singapore

[f] AI and HPC Research Center, IIT Bhubaneswar, India

Corresponding Author: Satish K Panda

First and last name: Satish K, Panda

E-mail address: skpanda@iitbbs.ac.in

**Abstract**

Optical Coherence Tomography (OCT) is an indispensable tool in ophthalmology, offering detailed cross-sectional images of the eye's anterior and posterior segments. However, the presence of speckle noise and other imaging artifacts can significantly degrade image quality, complicating the diagnosis of ophthalmic conditions. To address this challenge, we developed an enhanced denoising model based on a Residual U-Net architecture. This model was applied to both Anterior Segment OCT (ASOCT) and Posterior Segment OCT (PSOCT) images, aiming to reduce noise and improve overall image clarity while preserving essential anatomical features. Our model demonstrated considerable improvements in image quality metrics. For PSOCT images, the Peak Signal to Noise Ratio (PSNR) improved to $34.343 \pm 1.113$, and the Structural Similarity Index Measure (SSIM) reached $0.885 \pm 0.030$. For ASOCT images, the results were also significant, with PSNR at $23.525 \pm 0.872$ dB and SSIM at $0.407 \pm 0.044$. These enhancements were indicative of better preservation of tissue integrity and textural details. The implementation of the Residual U-Net model has proven effective in reducing noise and enhancing the diagnostic quality of OCT images across both ASOCT and PSOCT modalities. This dual functionality confirms the model's versatility and potential in clinical settings, contributing to more precise evaluations and potentially reducing the need for repeated imaging sessions.

# Statements and Declarations

**Funding**

This study was supported by SERB (SRG/2022/000922), IITI DRISHTI (CPS/CF/PG/2023/000007), ICMR (GRANTATHON2/MH-14/2024-NCD-II), and grants from the National Medical Research Council, Singapore (TA: NMRC/STAR/0023/2014 and MOH-00435).

**Competing Interests**

The authors have no conflicts of interest to declare.

**Author Contributions**

ANP (first author): Data preprocessing, network design and training, and manuscript preparation. JSG: Data preprocessing and manuscript preparation. RK, TA, TAT: Patient recruitment and data acquisition, and data labelling, Fund acquisition, SKP (corresponding author): Fund acquisition, method conceptualization, and manuscript preparation.

**Ethics Approval**

<u>Study approval statement</u>: "*This study protocol was reviewed and approved by Institute ethics committee of each participating institutes.*"

<u>Consent to participate statement</u>: *"Written informed consent was obtained for all subjects"*

**Data Availability**

The data used in this study will be made available on request.

**Introduction**

Optical Coherence Tomography (OCT) has become an indispensable imaging modality in ophthalmology, enabling high-resolution, cross-sectional visualization of the anterior and posterior segments of the eye [1]. This technology is crucial in diagnosing various ocular conditions, such as glaucoma and retinal degeneration [2]. However, the clinical efficacy of OCT is often compromised by intrinsic factors such as speckle noise, blood vessel shadows, and poor tissue visibility, which degrade image quality and pose challenges during diagnosis [3].

Historically, techniques like adaptive filtering and wavelet denoising have been employed to mitigate noise in OCT images [4], but these methods can blur crucial image features and often require manual tuning, making them less practical in fast-paced clinical settings. Such limitations have spurred interest in advanced solutions, particularly deep learning models, which promise greater efficacy through learned representations. The U-Net architecture, initially developed for biomedical image segmentation, has been notably effective due to its symmetric encoder-decoder structure, enhanced by skip connections that facilitate detailed context incorporation and precise localization [5].

Further advancements led to the integration of residual blocks within the U-Net architecture, resulting in the Residual U-Net, which enhances the learning dynamics and performance of complex noise patterns in OCT images [6]. This development aligns with emerging techniques in deep learning, such as self-supervised learning approaches, which have shown significant promise in noise reduction without requiring cleanly labeled data. For example, the Noise2Noise training strategy, which trains models directly on noisy targets, has been adapted for medical imaging and demonstrates substantial improvements in denoising without clean ground truth [7].

Despite these advancements, the application of Residual U-Nets to OCT image denoising faces many challenges, such as handling diverse noise types across different OCT devices and the dependency on large, annotated datasets for training, which limits the generalizability of the models [8]. In this context, emerging generative models like diffusion models offer a novel alternative [9]. These models progressively learn to reverse a diffusion process to reconstruct clean images from noisy inputs, accommodating various noise intensities and distributions without relying on extensive labeled datasets. This capability could potentially offer a scalable solution to noise reduction across diverse OCT systems. In this study, we propose a novel application of the Residual U-Net model, specifically designed to address the varieties of noise present in OCT images [10]. Our approach draws parallels with OCT-GAN, as proposed by Cheong et al. (2021), which aims for single-step shadow and noise removal from OCT images, focusing on specific noise and shadow artifacts inherent to OCT images [11]. While similar in intent, our study diverges by employing the Residual U-Net architecture with novel loss functions and metrics tailored for enhancing the denoising process, incorporating insights from recent studies on image quality assessment and deep learning architectures. [12-21].

**Methods**

**Data collection:**

A total of 100 ASOCT images and 2000 PSOCT images were obtained from the Singapore Eye Research Institute, Singapore, and Aravind Eye Care Systems, Madurai, India, respectively. Written informed consent was obtained for all subjects. The study was conducted following the tenets of the World Medical Associations Declaration of Helsinki and had ethics approval from the Institutional Review Board of each participating institute. A detailed description of OCT imaging and patient recruitment is described in our previous works [22-23].

**Database preparation:**

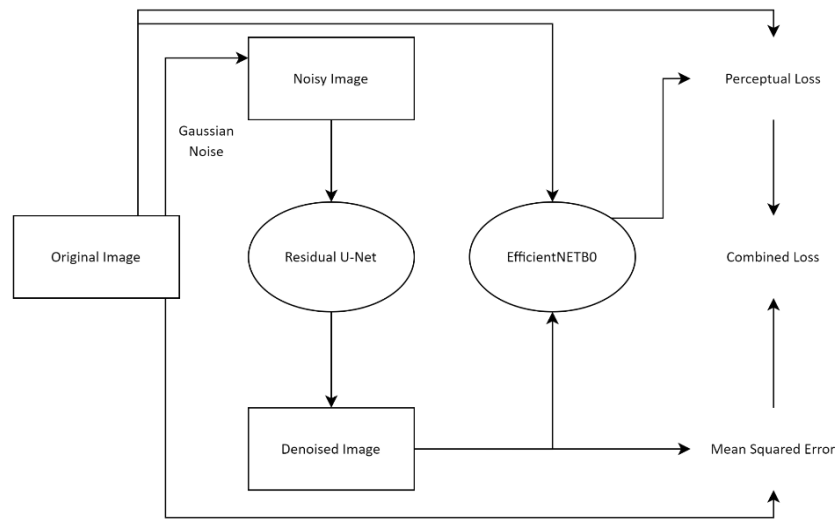

Figure 1: Architecture

We split the PSOCT images into training (70%), validation (15%), and test datasets (15%) and trained the proposed network on the training set images. Furthermore, all available ASOCT images were used exclusively for testing and were not part of the training dataset. To model realistic noise conditions, uniform Gaussian noise was introduced to the images during preprocessing. The denoising model centered on a Residual U-Net architecture, enhanced by incorporating a pre-trained network, i.e., EfficientNet, for feature extraction. This setup was coupled with a hybrid loss function that integrated Mean Squared Error (MSE) with additional perceptual quality components from the pre-trained network, aiming to optimize both the fidelity of image reconstruction and the perceptual integrity of the denoised outputs. Figure 1 represents the denoising process using a combination of Residual U-Net and EfficientNet architectures. Gaussian noise was added to an original image to simulate noisy conditions, which was then processed through the Residual U-Net for noise reduction. The resultant denoised image, alongside the original image, was evaluated using EfficientNet to calculate perceptual loss. This metric, combined with MSE, formed a custom loss function guiding the network's training.

**Model architecture: Residual U-Net**

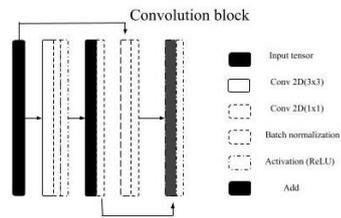 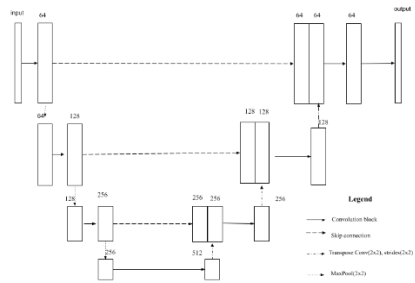

Figure 2(a): Convolution Block    Figure 2(b) Residual U-Net

Our study employed a Residual U-Net architecture, an advanced adaptation of the classic U-Net model enhanced with residual blocks for effective deep neural network training by mitigating the vanishing gradient problem. The model architecture began with an input layer designed to process images of size $200 \times 400$ accommodating the standardized preprocessed OCT images. Each convolutional block (shown in Fig. 2(a)) within the architecture consisted of two 3x3 convolutional layers, followed by batch normalization and ReLU activation layers to foster non-linear learning capabilities. Importantly, these blocks included residual connections where the input to each block was added to its output through a 1x1 convolution and batch normalization, enhancing feature propagation across the network.

The model featured a symmetric layout with a contracting path to capture context and an expansive path to enable precise localization (see Fig. 2(b)). Downsampling was achieved via MaxPooling2D layers, reducing feature map dimensions and increasing the receptive field, while upsampling employed Conv2DTranspose layers to restore spatial dimensions, aided by skip connections that reintroduced high-resolution features from earlier layers. This arrangement helped in maintaining the integrity of structural details in the images. The final output was generated through a 1x1 convolution followed by a sigmoid activation, ensuring the resulting denoised image would retain appropriate pixel value normalization.

**Noise modelling**

The preprocessing step involved converting each image into *float32* type followed by normalization to have pixel values between 0 and 1. The images were then resized or padded to a uniform size $200 \times 400$ to ensure compatibility with the input layer of our network. We applied a random rotation of $\pm 10$ degrees and a translation of ±10 pixels as a process of data augmentation, which was applied only to the training dataset. To enhance the model's generalization ability, we introduced variations through random noise addition. This involved Gaussian noise addition scaled by a random factor between 0.02 and 0.5 into the images. Such stochastic perturbations simulate realistic variations and noise conditions encountered in clinical environments, thereby preparing the model to handle diverse and unforeseen noise patterns effectively.

**Feature extraction**

To counter the blurring effects associated with MSE while using it as a sole loss function in image denoising tasks, our model integrated a perceptual loss component along with MSE, to enhance the model's ability to preserve high-level features and textures. These are crucial for maintaining the perceptual quality of the denoised images. Instead of directly applying MSE to the pixel values, our approach utilized a pre-trained model (i.e., EfficientNetB0) to extract intricate image features, making it a central component of our loss function.

The feature extraction process began with preprocessing both targets (true) and predicted images using the standard preprocessing method of EfficientNetB0 and aligning them with the input distribution as expected by the network. This step was essential for optimizing the network's capability for accurate feature extraction. Subsequently, we obtained feature maps from both sets of images using the EfficientNetB0 model. The perceptual loss was then computed by applying MSE between these high-dimensional feature representations rather than on the raw pixel data.

**Loss Functions**

In our model, the loss function played a critical role in optimizing the quality of denoised images. It comprised a weighted average MSE and a perceptual loss, each addressing distinct aspects of image fidelity.

*Mean squared error*

The MSE function is presented in Eq.1, where N, M, and C represent the number of rows, columns, and channels in the image, respectively. This formula computes the average squared difference between the actual pixel values ($I_{true,ijk}$) and the predicted pixel values ($I_{pred,ijk}$) providing a straightforward quantification of pixel-wise error. In the context of image denoising, MSE ensures that the denoised output closely matches the target image in terms of raw pixel values. This metric is beneficial for minimizing noise as it directly penalizes the model for discrepancies between the denoised image and the ground truth at the pixel level, contributing to the overall sharpness and accuracy of the reconstructed images. This loss can be computed using the following expression:

$$MSE = \frac{1}{N \times M \times C} \sum_{i=1}^{N} \sum_{j=1}^{M} \sum_{k=1}^{C} \left(I_{true,ijk} - I_{pred,ijk}\right)^2 \qquad (1)$$

where $i, j$, and $k$ represent the row index, column index, and channel index of the image, respectively.

*Perceptual loss*

While MSE is valuable for pixel accuracy, it does not inherently account for human perceptual differences or higher-level image features, such as textures and edges. These features are crucial for clinical assessments. Perceptual loss (PL) addresses this limitation by utilizing features extracted from a pre-trained deep-learning model. It measures the squared differences in high-level features extracted by a neural network from the true ($\phi(I_{true})$) and predicted images ($\phi(I_{pred})$) across multiple layers $l$. This loss emphasizes perceptual similarities, such as texture and edges, by averaging errors over feature maps $N_l$ in each layer. By focusing on these high-level features, perceptual loss helps preserve textural details and structural integrity, ensuring that the denoised images not only free of noise but also maintain fidelity to clinically relevant features. Eq. 2 illustrates the mathematical expression for this loss function

$$Perceptual\ Loss = \sum_{l=1}^{l} \frac{1}{N_l} \sum_{i=1}^{N_l} \left(\phi_l(I_{true})_i - \phi_l(I_{pred})_i\right)^2 \qquad (2)$$

*A weighted average loss function with MSE and PL*

We utilized a weighted combination of these loss functions to harness the strengths of both (see Eq. 3). It allowed the model to minimize pixel-level errors while maintaining the perceptual qualities of the OCT images and ensured that the denoising process would enhance both the quantitative and qualitative aspects of the image simultaneously.

$$combined\ loss = (1-\alpha) \times MSE(y_{true}, y_{pred}) + \alpha \times PL(y_{true}, y_{pred}) \qquad (3)$$

**Quantitative Metrics**

*Peak signal-to-noise ratio*

The Peak Signal-to-Noise Ratio (PSNR) is a widely used metric to evaluate the quality of reconstructed or denoised images by comparing the ratio of the maximum possible power of a signal to the power of corrupting noise that degrades its quality. Eq. 4 shows the mathematical expression for PSNR, where $f_O$ is the original image and $f_L$ is a denoised image. A higher PSNR value suggests a lower level of error or noise with better reconstruction, as it signifies that the denoised image is closer to the original with fewer deviations in pixel values.

$$PSNR = -10 \times log_{10} \frac{|f_O - f_L|^2}{|f_O|^2} \qquad (4)$$

*Structural similarity index measure*

In contrast to PSNR, the Structural Similarity Index Measure (SSIM) is an advanced metric used to assess the similarity between two images by considering their structural information, texture, and contrast. The mathematical expression for SSIM is given in Eq.5, where $\mu_x, \mu_y$ the average pixel values of two images, indicate the mean luminance while $\sigma_x^2, \sigma_y^2$ representing their variances reflecting the texture and detail richness. The covariance $\sigma_{xy}$ measures how changes in one image correspond with the other, capturing shared structural and contrast patterns. Constants $C_1$ and $C_2$ prevents division by zero and thus stabilizes the measurements under uniform conditions. The numerator in Eq. 5 emphasizes luminance and contrast correlations, while the denominator normalizes these values. It ensures that the SSIM metric reflects the perceptual similarities and dissimilarities accurately, making it highly relevant for evaluating the effectiveness of the image-processing techniques

$$SSIM(x, y) = \frac{(2\mu_x \mu_y + C_1)(2\sigma_{xy} + C_2)}{(\mu_X^2 + \mu_y^2 + C_1)(\sigma_x^2 + \sigma_y^2 + C_2)} \qquad (5)$$

*Interpretation of metrics*

Both PSNR and SSIM are crucial for a holistic evaluation of image denoising techniques. PSNR measures the error at a pixel level, indicating the magnitude of noise relative to signal strength. In contrast, SSIM offers insights into how the changes due to denoising affect perceived image quality, focusing on attributes that contribute to the overall visual appreciation. Higher scores in both metrics typically correlate with better perceptual quality, reflecting fewer artifacts, better feature preservation, and overall image fidelity. Together, these metrics enabled us to achieve a robust model performance in both quantitative and qualitative terms.

**Model Training:**

To effectively train the Residual U-Net network on OCT images, we utilized a comprehensive and methodical approach. Our training procedure was carried out over 300 epochs with a learning rate of 1e-4. We used the Adam optimizer due to its adaptive learning rate capabilities. It also helped us to achieve faster convergence during the training process. The loss function employed was a weighted combination of MSE and PL. The weight for the perceptual loss was set at 0.8, prioritizing the perceptual integrity of the denoised images.

**Results**

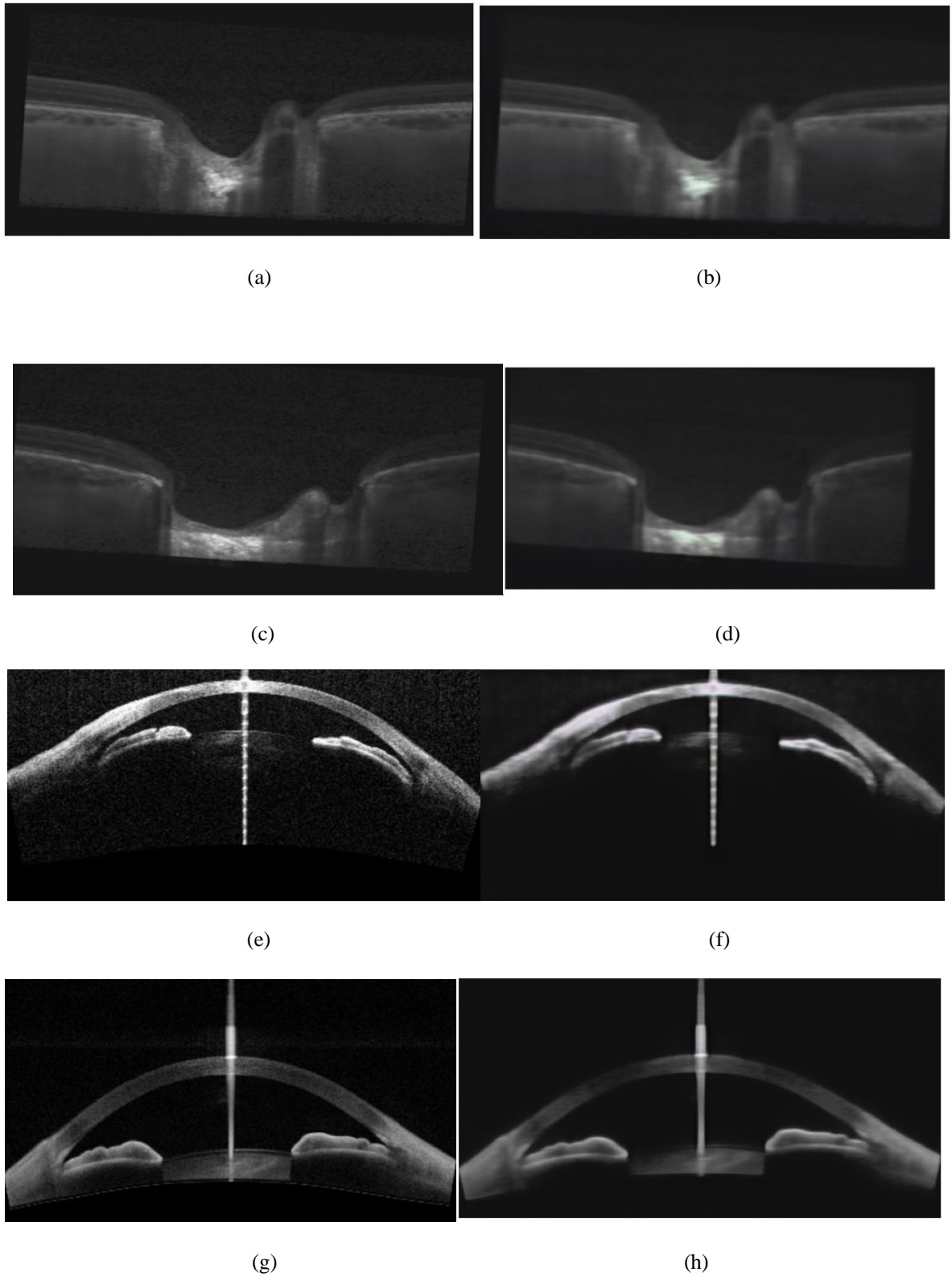

Figure 3 (a)-(h): Noisy and Corresponding Clean images of PSOCT and ASOCT Images

**Qualitative and quantitative comparison:** Our proposed approach effectively minimized the noise while preserving critical morphological features, such as the cornea and iris boundaries in ASOCT images and retinal and connective tissue boundaries in PSOCT images. Figure 3a and Figure 3c showcase raw PSOCT images characterized by prevalent speckle noise, which can mask critical anatomical features and diminish the images' diagnostic value. In contrast, the denoised images (Fig. 3(b), 3(d)) demonstrate a notable reduction in speckle noise, resulting in enhanced clarity and improved visibility of ocular structures. Figure 3e and Figure 3g display raw ASOCT images with inherent speckle noise. This noise can obscure finer anatomical details and degrade the diagnostic utility of the image. Conversely, the denoised images (Fig. 3(f), 3(h)) illustrate a significant reduction of speckle noise, leading to a clearer and more discernible depiction of the ocular structures.

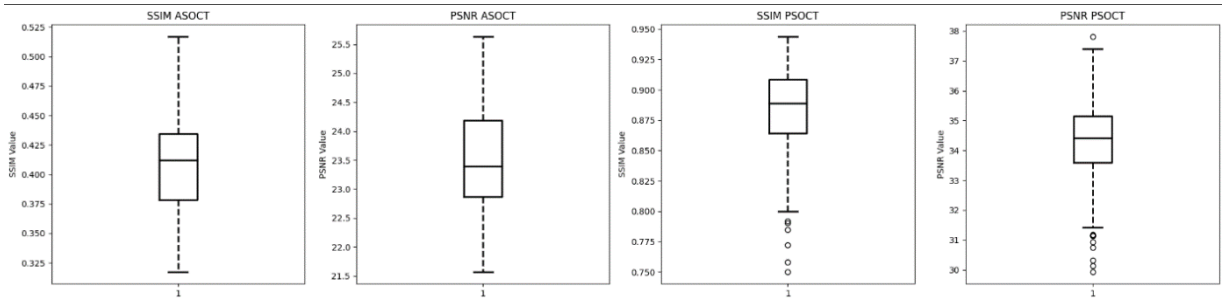

Figure 4: SSIM, PSNR for PSOCT and ASOCT Test Images

The quantitative assessment of image quality using SSIM and PSNR is provided in Figure 4 For ASOCT images, the SSIM was found to be 0.407 $\pm$ 0.044, and the PSNR was measured at 23.525 $\pm$ 0.872 dB. In the case of PSOCT images, the SSIM was found to be 0.885 $\pm$ 0.030, and the PSNR reached 34.343 $\pm$1.113 dB.

The metrics demonstrate enhanced performance and consistency of the proposed denoising algorithm, highlighting significant improvements in image quality and noise reduction.

**Discussion**

Our model, uniquely trained on a dataset of PSOCT images, demonstrated its capability to effectively remove noise when applied to ASOCT images without retraining. The quantitative results highlighted superior performance on PSOCT images with SSIM values reaching up to 0.900 and PSNR as high as 37, in contrast to ASOCT images which exhibited an SSIM around 0.400 and PSNR near 25. Qualitatively, the denoising process significantly reduced noise and preserved high levels of details and structural integrity in PSOCT images. However, in ASOCT images, we observed less clarity and more noise in deeper structures than in PSOCT. This showcases the robustness of our denoising approach and its efficacy across different types of OCT images, underscoring substantial improvements in image clarity and quality especially noted in PSOCT images.

The model's effectiveness across these diverse datasets can be attributed to several factors. Both ASOCT and PSOCT images share inherent characteristics due to the underlying OCT technology, including similar noise patterns and textures, allowing the model trained on PSOCT to generalize well to ASOCT. The architecture of the model, specifically the use of a Residual U-Net, effectively learns noise patterns which likely facilitated the superior performance on PSOCT images. Furthermore, the combined loss function integrating MSE with PL ensured pixel-level accuracy and enhanced perceptual quality, supporting high-quality denoising across different OCT datasets. This dual approach not only helped to preserve structural details after denoising but also improved visual clarity, making the model highly effective for clinical use where precision and perceptual clarity are paramount. This integrated strategy enhanced the model's adaptability and demonstrated its potential utility in diverse clinical settings.

Despite the effectiveness of our denoising model, several limitations persist, including variable performance across different OCT modalities, decreased effectiveness in extremely noisy images, and dependency on the diversity and quality of the training dataset. These challenges highlight the need for further refinement and development of the model to improve its generalizability and robustness. Looking ahead, the incorporation of diffusion models into our denoising framework appears to be a promising strategy. Known for their robust handling of complex noise distributions, diffusion models could significantly enhance the denoising capabilities of our approach, extending its applicability across broader medical imaging contexts and potentially setting new standards for image quality in both clinical and research environments.

**Figure Captions**

**Fig. 1** Overview of the proposed deep learning algorithm. This schematic illustrates the computational steps and data flow within our custom Residual U-Net model designed for OCT image denoising

**Fig. 2(a)** Residual U-Net Convolution block. Detail of a single convolution block used in our Residual U-Net model, showing the layers, activation functions, and skip connections

**Fig. 2(b)** Residual U-Net Architecture. Full architecture of the Residual U-Net, depicting how multiple convolution blocks are arranged and interconnected to enhance denoising performance

**Fig. 3(a)** Raw PSOCT image. Example of a raw Posterior Segment OCT image exhibiting typical speckle noise

**Fig. 3(b)** Clean and noise-free PSOCT image. Processed image corresponding to Fig. 3(a), demonstrating the effectiveness of our denoising model in enhancing image clarity

**Fig. 3(c)** Raw PSOCT image. Another example of a raw PSOCT image, showcasing initial image quality issues

**Fig. 3(d)** Clean and noise-free PSOCT image. Resulting image from Fig. 3(c) post-denoising, highlighting improved structural visibility

**Fig. 3(e)** Raw ASOCT image. Display of a raw Anterior Segment OCT image before application of the denoising

**Fig. 3(f)** Clean and noise-free ASOCT image. Processed image from Fig. 3(e), showing the reduction of noise and enhanced detail

**Fig. 3(g)** Raw ASOCT image. A second example of an unprocessed ASOCT image

**Fig. 3(h)** Clean and noise-free ASOCT image. The image from Fig. 3(g) after undergoing denoising, illustrating significant improvements in image quality

**Fig. 4** Box plots of SSIM and PSNR for ASOCT and PSOCT datasets. These plots show the statistical distribution of image quality metrics before and after denoising, indicating the efficacy of the model across different image types